# Spin Amplification by Controlled Symmetry Breaking for Spin-Based Logic

**Roland K Kawakami**
Department of Physics, The Ohio State University, Columbus, OH 43210

E-mail: kawakami.15@osu.edu

**Abstract.** Spin amplification is one of the most critical challenges for spintronics and spin-based logic in order to achieve spintronic circuits with fan-out. We propose a new concept for spin amplification that will allow a small spin current in a non-magnetic spin channel to control the magnetization of an attached ferromagnet. The key step is to bring the ferromagnet into an unstable symmetric state (USS), so that a small spin transfer torque from a small spin current can provide a magnetic bias to control the spontaneous symmetry breaking and select the final magnetization direction of the ferromagnet. Two proposed methods for achieving the USS configuration are voltage-controlled Curie temperature (VC-$T_C$) and voltage-controlled magnetic anisotropy (VC-MA). We believe the development of new 2D magnetic materials with greater tunability of VC-$T_C$ and VC-MA will be needed for practical applications. A successful realization of spin amplification by controlled symmetry breaking will be important for the implementation of existing spin-logic proposals (e.g. "All Spin Logic") and could inspire alternative ideas for spintronic circuits and devices.

The advent of spin channel materials with long spin diffusion lengths at room temperature (e.g. several microns for graphene [1-3]) enables prospects for spin-based logic relying on the flow of spin currents. The potential advantages of spin-based logic come from the integration of logic with nonvolatile ferromagnetic memory. This nonvolatility enables spintronic circuits with non-von-Neumann architectures, which may have enhanced performance beyond CMOS for data-intensive applications even if the speed or energy per operation of an individual device is not better than CMOS [4].

One of the primary challenges for spintronics and spin-based logic is to develop a process of spin amplification. Specifically, our usage of the term "spin amplification" refers to the concept of a small number of spins controlling a large number of spins, which will ultimately enable fan-out in spintronic circuits. Here, we present a method of spin amplification based on the control of spontaneous symmetry breaking in ferromagnets, which could make existing proposals such as All Spin Logic (ASL) [5] more feasible and may inspire alternative ideas for spintronic circuits. We specifically discuss two physical mechanisms to achieve spin amplification by controlled symmetry breaking that are consistent with low power dissipation: (1) voltage control of Curie temperature [6, 7] and (2) voltage control of magnetic anisotropy [8]. 2D materials are particularly well suited for this purpose because electrostatic gates can strongly tune the carrier density and thus control the magnetic properties in proposed magnetic transition metal dichalcogenides (TMDs) [9-12].

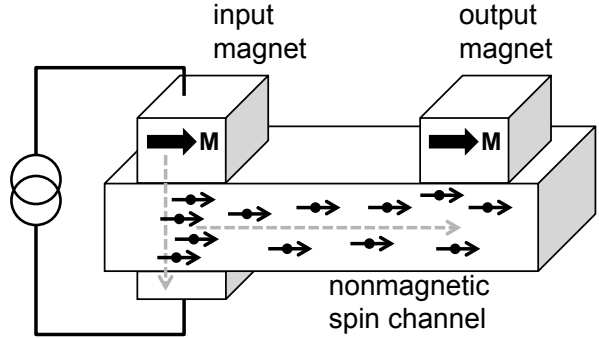

**Figure 1.** Relevant components of an All Spin Logic device. Charge current through the input magnet produces non-equilibrium spin polarization inside the nonmagnetic spin channel by electrical spin injection. These spin-polarized electrons transport to the output magnet as a pure spin current (electron diffusion without net charge flow). Finally, the spin current into the output magnet is meant to write the output magnetization via spin transfer torque.

For concreteness about our concept of spin amplification, it is worthwhile to discuss relevant parts of the ASL proposal. We consider a nonmagnetic spin channel with an input ferromagnet and an output ferromagnet as shown in Figure 1. By driving electric current from the input ferromagnet to the spin channel and exiting through a ground point, this generates electron spin polarization inside the spin channel (i.e. spin injection). Subsequently, the spin polarization diffuses through the spin channel via electron diffusion and reaches the output ferromagnet. These processes of spin injection and spin diffusion are well established in various spin channel materials including metals, semiconductors, and graphene [13-16]. The final and most critical step for ASL is to have the electron spin polarization in the spin channel write the magnetization state of the output ferromagnet via spin transfer torque. This is the step that defines our sense of spin amplification. Consider the electron spins in the channel as the "small number of spins" and the ferromagnet as the "large number of spins". Then this step of writing the magnetization of the output ferromagnet is a situation where a small number of spins is controlling a large number of spins. Our goal of spin amplification is to have a smaller number of spins in the channel control an increasingly larger number of spins in the output ferromagnet.

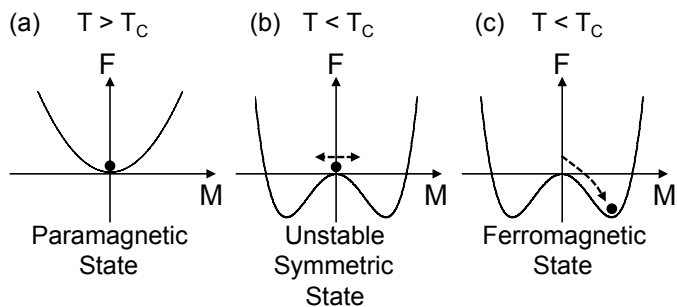

**Figure 2.** (a) Free energy (F) vs. Magnetization (M) for the output magnet in the paramagnetic state for temperature (T) above the Curie temperature ($T_C$), (b) For $T < T_C$, the zero magnetization state is an unstable equilibrium. We call this the "unstable symmetric state" because it maintains the ±M reflection symmetry of the F vs. M curve. (c) For $T < T_C$, the system reaches the ferromagnet ground state by spontaneously breaking the symmetry and choosing a particular sign for M.

The key idea is to write the output magnetization by controlling the "spontaneous symmetry breaking," the process by which a ferromagnet acquires its magnetization direction when it undergoes a phase transition from the paramagnetic state ($T>T_C$, Curie temperature) to the ferromagnetic state ($T<T_C$).

As illustrated in Figure 2, the free energy (F) vs. Magnetization (M) in the paramagnetic state has a minima at M=0 (Fig. 2a). As the temperature is lowered to below $T_C$, the F vs. M curve has a local maximum at M=0 and local minima at M≠0 (Fig. 2b). With the magnetization still at zero, the system is in the unstable symmetric state (USS). Due to this instability, the system will "spontaneously" choose a direction for the magnetization in order to lower the free energy (Fig. 2c). Theoretically, the choice only requires an infinitesimal magnetic bias, such as the Earth's magnetic field, stray magnetic fields, or random fluctuating magnetic fields. In principle, the magnetic bias could also come from spintronic effects such as spin transfer torque (STT), which we will use for the basis of spin amplification. In terms of the ASL device (Figure 1), we first get the output magnet into the unstable symmetric state (USS) in Figure 2b. Then we utilize spin transfer torque from the spin channel to provide the magnetic bias that will select the final magnetization state. In theory, the STT could be infinitesimally small, but in practice it needs be large enough to overcome any stray magnetic fields in the device. In any case, the amount of spin transfer torque needed should be orders of magnitude smaller than required in the original ASL proposal. Thus, by using a small STT to provide the magnetic bias to control the spontaneous symmetry breaking and write the output magnet, we are able to achieve our notion of spin amplification.

The method for getting the system into the USS configuration is important. Heating the output magnet to increase its temperature above $T_C$ is not a practical solution. This is likely to require too much power and the thermal process would be slow. A better option is to use electrostatic gates to control magnetic properties such as voltage control of the Curie temperature (VC-$T_C$) and voltage control of the magnetic anisotropy (VC-MA). In benchmarking studies, spin-logic devices based on voltage controlled switching tend to rate very well in terms of the energy per operation [17], so we anticipate that the use of VC-$T_C$ and VC-MA for spin amplification could also rate well for this metric.

The VC-$T_C$ was first observed in $In_{1-x}Mn_xAs$, a dilute magnetic semiconductor where Mn ions are coupled ferromagnetically by a hole-mediated exchange interaction [6]. By using a positive gate voltage to reduce the hole density, the exchange coupling between Mn ions is weakened and $T_C$ goes down. More recently, the VC-$T_C$ was achieved at room temperature in ultrathin cobalt films, where an increase in the electron density increases $T_C$ [7]. The likely explanation for the effect is a modulation of the magnetocrystalline anisotropy, which has a strong effect on $T_C$ in 2D ferromagnets [18]. In this regard, proposed 2D ferromagnetic TMDs are ideally suited for VC-$T_C$ due to the strong gate tunability of carrier density and highly anisotropic crystal structure [9-12].

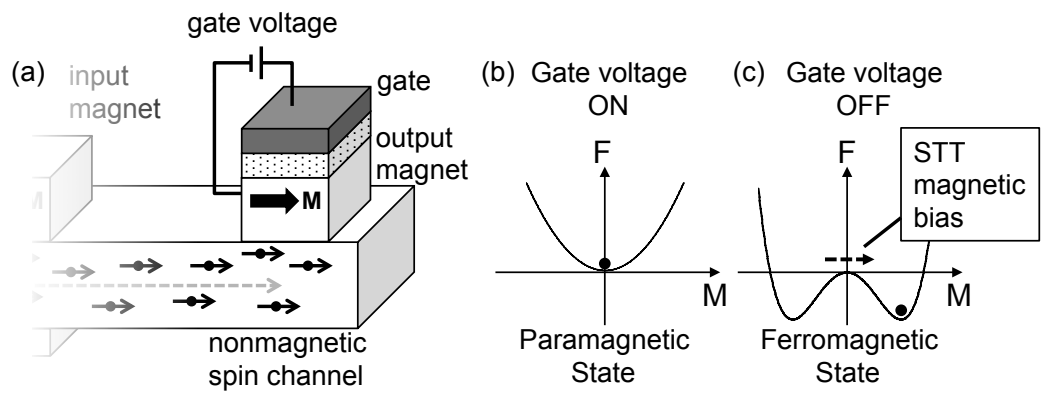

**Figure 3.** (a) Device for spin amplification by voltage control of Curie temperature (VC-$T_C$), (b) Free energy (F) vs. magnetization (M) with the gate voltage applied to move $T_C$ to below room temperature, (c) F vs. M with gate voltage removed to move $T_C$ to above room temperature. The direction of the final magnetization is determined by the STT magnetic bias.

The geometry for the VC-$T_C$ approach is shown in Figure 3a, with a gate dielectric and metal gate located on top of the output magnet. To write the magnetization state, the magnet is first brought into the paramagnetic state by applying the appropriate gate voltage ("ON"), which causes the magnetization to go to zero (Figure 3b). Next, spin current in the nonmagnetic spin channel is generated by electrical spin injection (or possibly spin pumping) from the input ferromagnet, which produces a magnetic bias on the output magnet via STT. The gate voltage is then released ("OFF") to place the output magnet in the USS configuration (see Fig. 2b), and the STT magnetic bias can determine the direction in which the magnetization will align to break symmetry. Figure 3c shows the final state of the output magnet in its ferromagnetic ground state, where the magnetization direction is determined by the STT magnetic bias controlling the spontaneous symmetry breaking.

The use of VC-MA provides an alternative method of reaching the USS configuration. The experimental demonstration of VC-MA in a solid state device was achieved in a (metal gate)/MgO/Fe structure, where a positive gate voltage reduces the strength of the perpendicular magnetic anisotropy of the Fe film [8]. Possible explanations for the change in magnetocrystalline anisotropy include (1) a change in the occupancy of *d*-orbitals [19] and (2) electric-field-induced modifications to the electronic band structure [20]. To illustrate the VC-MA writing process, we consider an output ferromagnet with an out-of-plane (z-axis) easy axis. To begin, a gate voltage is applied to turn the z-axis into the hard axis (i.e. x-y easy plane) and bring the magnetization in-plane ($M_Z=0$). Next, the gate voltage is released to turn the z-axis back into the easy axis and the system goes into a USS configuration with $M_Z=0$. From this configuration, the STT magnetic bias will push the system to the ferromagnetic ground state and determine the final direction of the output magnetization.

We also note that using the voltage controlled spin amplification (both VC-$T_C$ and VC-MA) has an inherent advantage of ensuring the directionality of information flow. For example, in a system consisting of multiple ferromagnets connected by a spin channel, each magnet has the ability to act as an input or an output. Operating as an input, the magnet generates spin polarization in the spin channel according to its magnetization direction through electrical spin injection (or possibly other methods such as microwave spin pumping or thermal spin injection). For electrical spin injection, a magnet is selected for this function simply by applying a current from the ferromagnet to the spin channel. However, once the spin polarization is inside the spin channel, it can diffuse in any direction and, in principle, write the magnetization of any of the other magnets. In such a situation, there is no control over directionality of the information flow. One approach to overcome this problem is to control the spin diffusion, such as in ASL where the contact resistances between the ferromagnet and spin channel are adjusted to promote directionality of the spin current. The beauty of our spin amplification method is that the spin diffusion does not have to be controlled because the output magnet is selected by applying a gate voltage. Thus, the directionality between an input magnet (selected by electrical spin injection current) and an output magnet (selected by gate voltage) is strictly enforced. This flexibility in the selection of input and output magnets to control the flow of information could potentially lead to completely new approaches for spin-based logic.

Finally, we will discuss how our proposed spin amplification scheme relates to previous device concepts and experimental work. In the initial ASL proposal, it was mentioned that perhaps a magnetic field pulse could be used to tip the magnetization of the output magnet into the USS configuration and have the STT select the output magnetization direction [5]. However, from an experimental point of view, the precision required for achieving the USS configuration on each output magnet seems unfeasible. However, these ideas certainly provide some inspiration for the concepts that we propose in this paper. We believe the approach using VC-$T_C$ or VC-MA is more feasible, although it will require a substantial dedicated effort in developing new 2D magnetic materials with greater tunability of $T_C$ and magnetic anisotropy. Experimentally, studies of STT from non-local spin current have been performed on both metal spin channels and graphene [21, 22]. In both cases, the amount of spin torque is small. For metals, researchers have achieved a switching of output magnetization in zero field, but channels have to be extremely short and the output magnetic moment has to be very small. For graphene, the reports of STT switching rely on having a very large magnetic field applied to bring the system to just below the switching threshold. This confirms that the STT effect in their devices is very small (i.e. unable to perform STT switching in zero magnetic field). Our proposed spin amplification will provide a way for such a small STT to produce magnetization switching in zero magnetic field.

In conclusion, we propose a new method for spin amplification based on controlled symmetry breaking that allows a spin current in a non-magnetic spin channel to write the magnetization of an attached output magnet. While the concepts could be applied to many material systems, it is particularly well-suited for spintronic devices based on 2D materials. Graphene is an ideal material for the spin channel because it has the longest spin diffusion length at room temperature of any material. Proposed magnetic TMDs appear to have the desired characteristics needed to achieve strong electric field tuning of Curie temperature and magnetic anisotropy. Developing such materials and achieving spin amplification will be an important advance for spin-based logic and 2D spintronics.

We acknowledge stimulating discussions with Ezekiel Johnston-Halperin, Roberto Myers, Chris Hammel, Beth Bushong, Steven Koester, Kaushik Roy, Paul Crowell, Michael Flatté, Justin Young, and Sara Mueller, and the support of ONR (N00014-14-1-0350), NSF-NRI (DMR-1124601), and C-SPIN, one of six centers of STARnet, a Semiconductor Research Corporation program sponsored by MARCO and DARPA.